\documentclass[traditabstract]{aa} 
\usepackage{graphicx}
\usepackage{txfonts}
\usepackage{natbib}
\bibpunct{(}{)}{;}{a}{}{,} 
\begin{document}
   \title{Inelastic Mg+H collision data for non-LTE applications in stellar atmospheres}

   \author{P. S. Barklem\inst{1}
          \and
          A. K. Belyaev\inst{1,2,3,4,5}
          \and
          A. Spielfiedel\inst{4}
          \and
          M. Guitou\inst{5}
          \and
          N. Feautrier\inst{4}
          }

   \institute{Department of Physics and Astronomy, Uppsala University, Box 515 S-75120 Uppsala, Sweden 
         \and
             Department of Theoretical Physics, Herzen University, St.\ Petersburg 191186, Russia
         \and
             Max-Planck-Institut f{\"u}r Astrophysik, Karl-Schwarzschild-Strasse 1, 857 41 Garching bei M{\"u}nchen, Germany
         \and
             LERMA, Observatoire de Paris, 92195 Meudon Cedex, France
         \and
             Universit\'e Paris-Est,  Laboratoire Mod\'elisation et Simulation Multi-Echelle, MSME UMR 8208 CNRS, 5 Bd Descartes, 
             77454 Marne-la-Vall\'ee, France   
             }

   \date{Received  21 February 2012 ; accepted 21 March 2012}

 
  \abstract
  {Rate coefficients for inelastic Mg+H collisions are calculated for all transitions between the lowest seven levels and the ionic state (charge transfer), namely Mg($3s^2 \: ^1\mathrm{S}$, $3s3p \: ^3\mathrm{P}$, $3s3p \: ^1\mathrm{P}$, $3s4s \: ^3\mathrm{S}$, $3s4s \: ^1\mathrm{S}$, $3s3d \: ^1\mathrm{D}$, $3s4p \: ^3\mathrm{P}$)+H(1s) and Mg$^+$($3s \: ^2\mathrm{S}$)+H$^-$. The rate coefficients are based on cross-sections from full quantum scattering calculations, which are themselves based on detailed quantum chemical calculations for the MgH molecule.  The data are needed for non-LTE applications in cool astrophysical environments, especially cool stellar atmospheres, and are presented for a temperature range of 500--8000~K.  From consideration of the sensitivity of the cross-sections to various uncertainties in the calculations, most importantly input quantum chemical data and the numerical accuracy of the scattering calculations, a measure of the possible uncertainties in the rate coefficients is estimated. }


   \maketitle
%

\section{Introduction}

The need for accurate data on inelastic atomic collisions involving hydrogen atoms for non-LTE line formation calculations in cool star atmospheres has been well documented \citep[e.g.][]{1984A&A...130..319S, 1993PhST...47..186L, 2001NewAR..45..559K, 2005ARA&A..43..481A}.   Non-LTE models have often, for lack of a better alternative, been forced to use the so-called ``Drawin formula''  \citep[see][]{1984A&A...130..319S,1968ZPhy..211..404D,1973PhLA...43..333D}, which has been shown to not contain the relevant physics nor provide order-of-magnitude estimates of the collision rates \citep{2011A&A...530A..94B}.  

Full quantum scattering calculations based on quantum chemical data have been done for Li+H \citep{2003PhRvA..68f2703B} and Na+H \citep{2010PhRvA..81c2706B}, and rate coefficients provided for astrophysical applications \citep{2003A&A...409L...1B, 2010A&A...519A..20B}.  The data have been applied to non-LTE calculations by \cite{2003A&A...409L...1B} and \cite{2009A&A...503..541L} in the case of Li, and \cite{2011A&A...528A.103L} in the case of Na.  In both cases, it has been found that direct excitation and deexcitation processes, $\mathrm{X}(nl) + \mathrm{H} \rightleftharpoons \mathrm{X}(n^\prime l^\prime) + \mathrm{H}$ where X is the atom of interest, are of little consequence for the statistical equilibrium.  However, charge transfer processes, $\mathrm{X}(nl) + \mathrm{H} \rightleftharpoons \mathrm{X}^+ + \mathrm{H}^-$, particularly where $nl$ corresponds to the first excited $S$-state, are important.  The resulting effects on abundance corrections in solar-type stars were often small, of order a few 0.01~dex. In giants the influence was larger, of order 0.05~dex, and even larger still in metal-poor stars reaching of order 0.1--0.2~dex.  Thus, these processes are important for accurate absolute abundances, and also for accurate relative abundances among dissimilar stars.

Alkalis have naturally been the first cases studied since they have only a single valence electron and thus present the easiest cases for quantum chemical calculations.  Also, Li and Na are of significant astrophysical interest.  We now present data for Mg, the motivation for which is two-fold.  First, it is of interest to study hydrogen collisions in more complex atoms and investigate the similarities and differences to the cases of alkalis: an atom with two valence electrons is the next logical step.  Second, Mg is of significant astrophysical importance.  Mg creates some of the strongest absorption lines in stellar spectra and is relatively easily detected even in low-quality spectra or in metal-poor stars. It is a tracer of the star formation history and chemical evolution of the galaxy since it is thought to be formed predominantly in supernovae resulting from massive stars.  In fact, Fe is often used as reference element in galactic chemical evolution studies since strong lines of Fe are ubiquitous in the visual spectra of cool stars. However, a number of different astrophysical sites are expected to have contributed significantly to the production of Fe and thus its history is quite complicated.  Given its relatively simple formation history, Mg is now a commonly used alternative reference element \citep[e.g.][]{2004A&A...416.1117C}.  We note that, to our knowledge, there are no previous theoretical or experimental studies of Mg+H inelastic collisions.

Quantum chemical calculations of the potentials and couplings have now been done for MgH including excited states \citep{2010CPL...488..145G}.  These were then used in full quantum scattering calculations, initially testing on a small number of states \citep{2011JPhB...44c5202G}, and later applying them to the seven lowest states of symmetry $^2\Sigma^+$ plus the ionic channel \citep{mgh_pra}.  This allows cross-sections to be calculated for direct excitation between the seven lowest states of Mg and for charge transfer processes involving these seven states.  These calculations should be sufficient for non-LTE line formation modelling of spectral lines corresponding to transitions between low-lying levels in cool stellar atmospheres.   Such models require rate coefficients and these are presented here.

\section{Collision rate coefficients}

The rate coefficients, $\langle \sigma \upsilon \rangle$, for excitation and deexcitation processes,
\begin{equation}
\mathrm{Mg}(3s \: nl \; ^{2S+1}L)+\mathrm{H}(1s) \rightleftharpoons \mathrm{Mg}(3s \: n^\prime l^\prime \; ^{2S^\prime +1}L^\prime)+\mathrm{H}(1s) ,
\end{equation}
and for the charge transfer processes, ion-pair production and mutual-neutralisation, involving the ionic state, 
\begin{equation}
\mathrm{Mg}(3s \: nl \; ^{2S+1}L) + \mathrm{H}(1s) \rightleftharpoons \mathrm{Mg}^+(3s \; ^2\mathrm{S})+\mathrm{H}^- ,
\end{equation}
are presented in Table~\ref{tab:rates}.  The coefficients have been obtained by folding the cross-sections with a Maxwellian velocity distribution from thresholds to 17 eV in total energy (corresponding to at least 10 eV in collision energy), and are presented for temperatures in the range 500--8000~K.  

\begin{table*}
\caption{Rate coefficients $\langle \sigma \upsilon \rangle$, in units of cm$^3$~s$^{-1}$, for selected temperatures in the range $T=500$--8000 K, for the excitation and deexcitation processes Mg($3s \: nl \; ^{2S+1}L$)+H($1s$)$\rightarrow$Mg($3s \: n^\prime l^\prime \; ^{2S^\prime +1}L^\prime$)+H($1s$), or where indicated ion-pair production Mg($3s \: nl \; ^{2S+1}L$)+H($1s$)$\rightarrow$Mg$^+$($3s \; ^2\mathrm{S}$)+H$^-$ and mutual neutralisation Mg$^+$($3s \; ^2\mathrm{S}$)+H$^-\rightarrow$ Mg($3s \: n^\prime l^\prime \; ^{2S^\prime +1}L^\prime$)+H($1s$).} 
\label{tab:rates}
\tiny
\begin{center}
\begin{tabular}{lcccccccc}
\hline
Initial & \multicolumn{8}{c}{final state}  \\
state  & $3s^2 \: ^1\mathrm{S}$ & $3s3p \: ^3\mathrm{P}^\mathrm{o}$ & $3s3p \: ^1\mathrm{P}^\mathrm{o}$ & $3s4s \: ^3\mathrm{S}$ & $3s4s \: ^1\mathrm{S}$ & $3s3d \: ^1\mathrm{D}$ &  $3s4p \: ^3\mathrm{P}^\mathrm{o}$&  Mg$^+$+H$^-$  \\
\hline
\multicolumn{9}{c}{\underline{ 500K}} \\
   $3s^2 \: ^1\mathrm{S}$  & ---  & $ 1.18$E$-41$  & $ 1.92$E$-58$  & $ 5.29$E$-65$  & $ 5.48$E$-68$  & $ 3.76$E$-71$  & $ 4.76$E$-73$  & $ 8.68$E$-83$ \\
  $3s3p \: ^3\mathrm{P}^\mathrm{o}$ & $ 2.95$E$-15$  & ---  & $ 7.84$E$-28$  & $ 8.81$E$-35$  & $ 5.56$E$-38$  & $ 2.99$E$-41$  & $ 2.81$E$-43$  & $ 1.37$E$-53$ \\
  $3s3p \: ^1\mathrm{P}^\mathrm{o}$ & $ 1.10$E$-14$  & $ 1.79$E$-10$  & ---  & $ 1.83$E$-17$  & $ 9.51$E$-21$  & $ 1.38$E$-24$  & $ 2.05$E$-26$  & $ 3.07$E$-35$ \\
  $3s4s \: ^3\mathrm{S}$   & $ 9.71$E$-14$  & $ 6.45$E$-10$  & $ 5.85$E$-10$  & ---  & $ 4.33$E$-13$  & $ 4.98$E$-17$  & $ 2.52$E$-18$  & $ 6.05$E$-27$ \\
  $3s4s \: ^1\mathrm{S}$ & $ 2.50$E$-13$  & $ 1.01$E$-09$  & $ 7.57$E$-10$  & $ 1.08$E$-09$  & ---  & $ 1.31$E$-12$  & $ 1.64$E$-15$  & $ 1.99$E$-22$ \\
   $3s3d \: ^1\mathrm{D}$  & $ 7.20$E$-14$  & $ 2.28$E$-10$  & $ 4.59$E$-11$  & $ 5.19$E$-11$  & $ 5.49$E$-10$  & ---  & $ 4.27$E$-12$  & $ 3.41$E$-20$ \\
  $3s4p \: ^3\mathrm{P}^\mathrm{o}$ & $ 3.19$E$-14$  & $ 7.52$E$-11$  & $ 2.41$E$-11$  & $ 9.22$E$-11$  & $ 2.41$E$-11$  & $ 1.50$E$-10$  & ---  & $ 3.29$E$-19$ \\
H$^-$+Mg$^+$ & $ 2.45$E$-13$  & $ 1.55$E$-10$  & $ 1.51$E$-09$  & $ 9.31$E$-09$  & $ 1.23$E$-07$  & $ 5.03$E$-08$  & $ 1.38$E$-08$  & --- \\
\multicolumn{9}{c}{\underline{2000K}} \\
    $3s^2 \: ^1\mathrm{S}$ & ---  & $ 3.49$E$-21$  & $ 2.48$E$-25$  & $ 2.20$E$-26$  & $ 4.07$E$-27$  & $ 4.72$E$-28$  & $ 1.96$E$-28$  & $ 9.23$E$-31$ \\
  $3s3p \: ^3\mathrm{P}^\mathrm{o}$ & $ 2.67$E$-15$  & ---  & $ 2.57$E$-15$  & $ 1.04$E$-16$  & $ 1.20$E$-17$  & $ 1.05$E$-18$  & $ 3.44$E$-19$  & $ 6.15$E$-22$ \\
  $3s3p \: ^1\mathrm{P}^\mathrm{o}$ & $ 9.46$E$-15$  & $ 1.28$E$-10$  & ---  & $ 6.58$E$-12$  & $ 4.97$E$-13$  & $ 2.09$E$-14$  & $ 8.84$E$-15$  & $ 1.30$E$-16$ \\
    $3s4s \: ^3\mathrm{S}$ & $ 6.33$E$-14$  & $ 3.89$E$-10$  & $ 4.95$E$-10$  & ---  & $ 6.26$E$-11$  & $ 1.89$E$-12$  & $ 1.98$E$-12$  & $ 6.68$E$-14$ \\
    $3s4s \: ^1\mathrm{S}$ & $ 1.88$E$-13$  & $ 7.23$E$-10$  & $ 6.02$E$-10$  & $ 1.01$E$-09$  & ---  & $ 6.27$E$-10$  & $ 2.69$E$-11$  & $ 1.52$E$-11$ \\
    $3s3d \: ^1\mathrm{D}$ & $ 2.95$E$-14$  & $ 8.61$E$-11$  & $ 3.42$E$-11$  & $ 4.11$E$-11$  & $ 8.48$E$-10$  & ---  & $ 2.03$E$-10$  & $ 8.16$E$-12$ \\
  $3s4p \: ^3\mathrm{P}^\mathrm{o}$ & $ 1.92$E$-14$  & $ 4.41$E$-11$  & $ 2.27$E$-11$  & $ 6.76$E$-11$  & $ 5.70$E$-11$  & $ 3.18$E$-10$  & ---  & $ 3.61$E$-12$ \\
H$^-$+Mg$^+$ & $ 2.13$E$-13$  & $ 1.85$E$-10$  & $ 7.88$E$-10$  & $ 5.36$E$-09$  & $ 7.60$E$-08$  & $ 3.01$E$-08$  & $ 8.50$E$-09$  & --- \\
\multicolumn{9}{c}{\underline{4000K}} \\
    $3s^2 \: ^1\mathrm{S}$ & ---  & $ 1.67$E$-17$  & $ 9.32$E$-20$  & $ 5.37$E$-20$  & $ 2.14$E$-20$  & $ 6.31$E$-21$  & $ 4.69$E$-21$  & $ 5.05$E$-22$ \\
  $3s3p \: ^3\mathrm{P}^\mathrm{o}$ & $ 4.87$E$-15$  & ---  & $ 2.76$E$-13$  & $ 7.95$E$-14$  & $ 2.07$E$-14$  & $ 4.35$E$-15$  & $ 2.68$E$-15$  & $ 1.47$E$-16$ \\
  $3s3p \: ^1\mathrm{P}^\mathrm{o}$ & $ 1.05$E$-14$  & $ 1.07$E$-10$  & ---  & $ 5.21$E$-11$  & $ 7.88$E$-12$  & $ 9.96$E$-13$  & $ 6.66$E$-13$  & $ 1.84$E$-13$ \\
    $3s4s \: ^3\mathrm{S}$ & $ 5.26$E$-14$  & $ 2.67$E$-10$  & $ 4.52$E$-10$  & ---  & $ 1.38$E$-10$  & $ 1.18$E$-11$  & $ 1.87$E$-11$  & $ 9.14$E$-12$ \\
    $3s4s \: ^1\mathrm{S}$ & $ 1.46$E$-13$  & $ 4.83$E$-10$  & $ 4.75$E$-10$  & $ 9.56$E$-10$  & ---  & $ 1.42$E$-09$  & $ 1.52$E$-10$  & $ 8.64$E$-10$ \\
    $3s3d \: ^1\mathrm{D}$ & $ 2.23$E$-14$  & $ 5.28$E$-11$  & $ 3.12$E$-11$  & $ 4.28$E$-11$  & $ 7.41$E$-10$  & ---  & $ 4.62$E$-10$  & $ 1.73$E$-10$ \\
  $3s4p \: ^3\mathrm{P}^\mathrm{o}$ & $ 1.55$E$-14$  & $ 3.03$E$-11$  & $ 1.95$E$-11$  & $ 6.31$E$-11$  & $ 7.38$E$-11$  & $ 4.31$E$-10$  & ---  & $ 4.59$E$-11$ \\
H$^-$+Mg$^+$ & $ 2.42$E$-13$  & $ 2.42$E$-10$  & $ 7.84$E$-10$  & $ 4.48$E$-09$  & $ 6.10$E$-08$  & $ 2.35$E$-08$  & $ 6.67$E$-09$  & --- \\
\multicolumn{9}{c}{\underline{6000K}} \\
    $3s^2 \: ^1\mathrm{S}$ & ---  & $ 4.52$E$-16$  & $ 8.22$E$-18$  & $ 7.89$E$-18$  & $ 3.96$E$-18$  & $ 1.64$E$-18$  & $ 1.34$E$-18$  & $ 4.56$E$-19$ \\
  $3s3p \: ^3\mathrm{P}^\mathrm{o}$ & $ 9.56$E$-15$  & ---  & $ 1.29$E$-12$  & $ 6.74$E$-13$  & $ 2.21$E$-13$  & $ 6.57$E$-14$  & $ 4.64$E$-14$  & $ 9.22$E$-15$ \\
  $3s3p \: ^1\mathrm{P}^\mathrm{o}$ & $ 1.33$E$-14$  & $ 9.86$E$-11$  & ---  & $ 1.10$E$-10$  & $ 1.89$E$-11$  & $ 3.82$E$-12$  & $ 2.70$E$-12$  & $ 2.34$E$-12$ \\
    $3s4s \: ^3\mathrm{S}$ & $ 5.39$E$-14$  & $ 2.18$E$-10$  & $ 4.63$E$-10$  & ---  & $ 1.81$E$-10$  & $ 2.66$E$-11$  & $ 4.03$E$-11$  & $ 4.87$E$-11$ \\
    $3s4s \: ^1\mathrm{S}$ & $ 1.42$E$-13$  & $ 3.76$E$-10$  & $ 4.19$E$-10$  & $ 9.50$E$-10$  & ---  & $ 1.73$E$-09$  & $ 2.83$E$-10$  & $ 3.20$E$-09$ \\
    $3s3d \: ^1\mathrm{D}$ & $ 2.23$E$-14$  & $ 4.22$E$-11$  & $ 3.20$E$-11$  & $ 5.29$E$-11$  & $ 6.56$E$-10$  & ---  & $ 6.35$E$-10$  & $ 4.60$E$-10$ \\
  $3s4p \: ^3\mathrm{P}^\mathrm{o}$ & $ 1.43$E$-14$  & $ 2.34$E$-11$  & $ 1.78$E$-11$  & $ 6.29$E$-11$  & $ 8.40$E$-11$  & $ 4.98$E$-10$  & ---  & $ 1.02$E$-10$ \\
H$^-$+Mg$^+$ & $ 2.80$E$-13$  & $ 2.67$E$-10$  & $ 8.88$E$-10$  & $ 4.37$E$-09$  & $ 5.47$E$-08$  & $ 2.08$E$-08$  & $ 5.89$E$-09$  & --- \\
\multicolumn{9}{c}{\underline{8000K}} \\
    $3s^2 \: ^1\mathrm{S}$ & ---  & $ 3.14$E$-15$  & $ 9.03$E$-17$  & $ 1.02$E$-16$  & $ 6.16$E$-17$  & $ 2.80$E$-17$  & $ 2.33$E$-17$  & $ 1.45$E$-17$ \\
  $3s3p \: ^3\mathrm{P}^\mathrm{o}$ & $ 1.79$E$-14$  & ---  & $ 2.79$E$-12$  & $ 1.91$E$-12$  & $ 7.16$E$-13$  & $ 2.49$E$-13$  & $ 1.79$E$-13$  & $ 7.26$E$-14$ \\
  $3s3p \: ^1\mathrm{P}^\mathrm{o}$ & $ 1.75$E$-14$  & $ 9.51$E$-11$  & ---  & $ 1.71$E$-10$  & $ 2.97$E$-11$  & $ 8.02$E$-12$  & $ 5.53$E$-12$  & $ 8.66$E$-12$ \\
    $3s4s \: ^3\mathrm{S}$ & $ 5.81$E$-14$  & $ 1.92$E$-10$  & $ 5.02$E$-10$  & ---  & $ 2.14$E$-10$  & $ 4.41$E$-11$  & $ 6.04$E$-11$  & $ 1.17$E$-10$ \\
    $3s4s \: ^1\mathrm{S}$ & $ 1.61$E$-13$  & $ 3.28$E$-10$  & $ 3.99$E$-10$  & $ 9.77$E$-10$  & ---  & $ 1.86$E$-09$  & $ 3.93$E$-10$  & $ 6.08$E$-09$ \\
    $3s3d \: ^1\mathrm{D}$ & $ 2.36$E$-14$  & $ 3.68$E$-11$  & $ 3.48$E$-11$  & $ 6.50$E$-11$  & $ 6.00$E$-10$  & ---  & $ 7.61$E$-10$  & $ 7.40$E$-10$ \\
  $3s4p \: ^3\mathrm{P}^\mathrm{o}$ & $ 1.41$E$-14$  & $ 1.90$E$-11$  & $ 1.73$E$-11$  & $ 6.41$E$-11$  & $ 9.13$E$-11$  & $ 5.48$E$-10$  & ---  & $ 1.51$E$-10$ \\
H$^-$+Mg$^+$ & $ 3.18$E$-13$  & $ 2.79$E$-10$  & $ 9.79$E$-10$  & $ 4.48$E$-09$  & $ 5.11$E$-08$  & $ 1.93$E$-08$  & $ 5.47$E$-09$  & --- \\
\hline
\end{tabular}
\end{center}
\end{table*}

The accuracy of the rate coefficients is determined by the accuracy of the cross-sections near the threshold.   As discussed in \cite{2010A&A...519A..20B}, this accuracy depends on a number of aspects, most importantly the accuracy of the input quantum chemical data (potentials and couplings), the possible contribution of neglected molecular symmetries and states, and the numerical accuracy of quantum scattering calculations.  As in \cite{2010A&A...519A..20B}, we have considered these points in detail for each transition and attempted to estimate ``fluctuation factors'' for each transition, which should be indicators of the uncertainty in the rate coefficients.  The probable range for a given rate coefficient can be estimated to be from the value given in Table~\ref{tab:rates} multiplied by the minimum value of the fluctuation factor, up to the same value multiplied by the maximum value of the fluctuation factor.  Table~\ref{tab:errors} presents minimum and maximum fluctuation factors for the rate coefficients in Table~\ref{tab:rates}.  

The main considerations used to determine these fluctuation factors are as follows.  Based on comparison of calculations using different quantum chemical data sets from different calculations done by us, we estimate the error from this source at of order 10\%.  Calculations were also done using two different programs for the quantum scattering calculations and the differences did not exceed 10\%.  Together, this results in minimum and maximum fluctuation factors of 0.8 and 1.2, respectively, for the majority of transitions.  However, additional or larger errors are likely in some cases.  It was shown in \cite{2011JPhB...44c5202G} that in the case of transitions between the lowest-lying states, inclusion of other molecular states and symmetries did not significantly affect the cross-sections; this is further discussed in \cite{mgh_pra}.  However, for higher-lying states, the cross-sections may be affected (most often increased) and this has been investigated via model estimates and included in the fluctuation factors.  In addition, there are reasons to believe that in some particular cases the quantum chemical data are more uncertain (e.g. for the highest-lying states) or that the cross-sections are more sensitive to uncertainties in the quantum chemical data  (e.g. the $3s^2 \: ^1\mathrm{S} \rightarrow 3s3p \: ^3\mathrm{P}^\mathrm{o}$ transition due to large splitting between adiabatic potentials at the avoided crossing).  In these cases, estimates of the additional uncertainty are made and included in the fluctuation factors.  

The fluctuation factors would be expected to have some variation with temperature since expected errors in the cross sections vary with collision energy.  Generally, we expect larger errors at lower temperatures where the near-threshold cross-sections have the largest contributions.  However, we have provided a single \emph{maximum} value for all temperatures, since our estimated fluctuation factors are generally small and their variation with temperature would be even smaller, and it is not clear that any estimates at this level would be meaningful.

It is worth commenting that the fluctuation factors here are significantly lower than those estimated for Na+H in \cite{2010A&A...519A..20B}.   The larger values for Na+H were predominantly due to larger disagreement found between results from different input quantum chemical data and results from different quantum scattering codes.  We have found no evidence for such large uncertainties in Mg+H, noting that the data and codes used here are all quite recent, while those compared in Na+H included older data and codes and a wider range of methods.

\begin{table*}
\caption{Estimated values of the ``fluctuation factors'' (a measure of the uncertainty) in the rate coefficients, given as factors of the rate coefficients in Table~\ref{tab:rates}.  Numbers above the main diagonal are the maximum values of the fluctuation factor for each transition, while numbers below the main diagonal are the minimum values for the same transition.} 
\label{tab:errors}
\tiny
\begin{center}
\begin{tabular}{lcccccccc}
\hline
Initial/final & \multicolumn{8}{c}{initial/final state}  \\
state   & $3s^2 \: ^1\mathrm{S}$ & $3s3p \: ^3\mathrm{P}^\mathrm{o}$ & $3s3p \: ^1\mathrm{P}^\mathrm{o}$ & $3s4s \: ^3\mathrm{S}$ & $3s4s \: ^1\mathrm{S}$ & $3s3d \: ^1\mathrm{D}$ &  $3s4p \: ^3\mathrm{P}^\mathrm{o}$&  Mg$^+$+H$^-$  \\
\hline
$3s^2 \: ^1\mathrm{S}$            & --- & 1.5 & 1.2 & 1.2 & 1.2 & 1.2 & 1.3 & 1.2  \\
$3s3p \: ^3\mathrm{P}^\mathrm{o}$ & 0.6 & --- & 1.2 & 1.2 & 1.2 & 1.2 & 1.3 & 1.2  \\  
$3s3p \: ^1\mathrm{P}^\mathrm{o}$ & 0.8 & 0.8 & --- & 1.4 & 1.2 & 2 & 1.3 & 2  \\
$3s4s \: ^3\mathrm{S}$            & 0.8 & 0.8 & 0.8 & --- & 1.2 & 2 & 1.3 & 1.2  \\
$3s4s \: ^1\mathrm{S}$            & 0.8 & 0.8 & 0.8 & 0.8 & --- & 1.2 & 1.4 & 1.2  \\
$3s3d \: ^1\mathrm{D}$            & 0.8 & 0.8 & 0.8 & 0.8 & 0.8 & --- & 3 & 5  \\
$3s4p \: ^3\mathrm{P}^\mathrm{o}$ & 0.4 & 0.4 & 0.4 & 0.4 & 0.4 & 0.3 & --- & 6  \\
Mg$^+$+H$^-$                      & 0.8 & 0.8 & 0.8 & 0.8 & 0.8 & 0.2 & 0.2 & ---  \\  
\hline
\end{tabular}
\end{center}
\end{table*}

\section{Discussion}

\begin{figure}
   \centering
   \includegraphics[width=90mm]{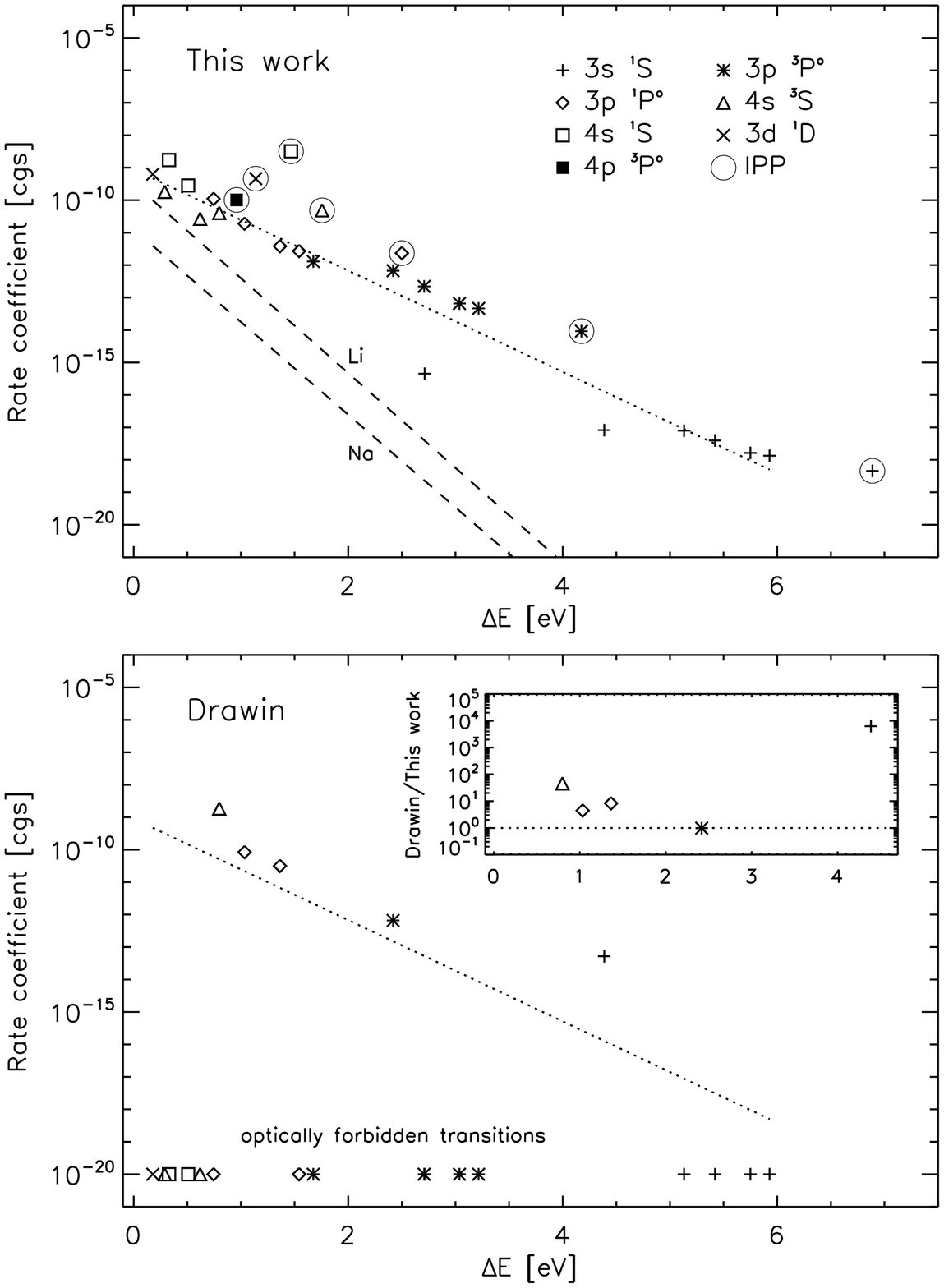}
      \caption{Rate coefficients at 6000~K for excitation and ion-pair production processes in Mg+H collisions, plotted against the transition energy $\Delta E$. Different symbols are used to denote the initial state of the transition following the key given in the upper panel. The upper panel shows the results from this work based on detailed quantum scattering calculations \citep{mgh_pra}.  Symbols inside circles refer to ion-pair production processes (IPP); i.e.\ the final state is the ionic state Mg$^+$($3s \; ^2\mathrm{S}$)+H$^-$.  The dotted line shows a linear fit to the data for excitation processes (i.e.\ excluding ion-pair production), which is repeated in the lower panel to aid comparison.  The dashed lines show the fits to the corresponding data for Li and Na from \cite{2011A&A...530A..94B}.
      The lower panel shows the results of the Drawin formula, and the inset shows the ratio with the quantum scattering calculations.  Optically forbidden transitions, where the Drawin formula is not applicable, are shown in the bottom of the panel.  
              }
         \label{fig:rates}
   \end{figure}

The rate coefficients for Mg+H collisions are plotted against transition energy, $\Delta E$, in Fig.~\ref{fig:rates}.  These plots are similar to those presented for Li+H and Na+H in \cite{2011A&A...530A..94B}, and include rate coefficients from quantum scattering calculations (upper panel) and those predicted by the commonly used Drawin formula (lower panel).  Rate coefficients for charge transfer processes (ion-pair production) are shown in addition to those for excitation.  A fit to the data for excitation processes is shown (dotted line), and the fits for Li+H and Na+H from \cite{2011A&A...530A..94B} are shown for comparison (dashed lines).  We emphasise that the actual data have a large scatter around these fits, and their purpose is to indicate the general behaviour, not to provide fits for use in models.

The most notable aspect of the Drawin formula results is that of the 21 excitation transitions considered, only 5 are optically allowed and can be calculated with the Drawin formula.  This is due to the fact that Mg has two valence electrons leading to the presence of singlet and triplet spin terms, between which optical transitions are forbidden.   As seen in Li and Na, there is a tendency for the Drawin results to be generally larger than those from quantum scattering calculations by a few orders of magnitude with a significant variation, here ranging from zero to four orders of magnitude larger.

Some general features of the rate coefficients for excitation from quantum mechanical calculations are very similar to those seen in Li and Na.  As expected, there is a correlation with the transition energy, although with a scatter of a few orders of magnitude, since the threshold energy is a main parameter in determining the rate coefficient at low temperatures.  There are also strong secondary correlations among transitions from a specific initial state.  

On the other hand, comparison of the general magnitude of the rate coefficients finds significant differences compared to Li and Na.  We see that at the same $\Delta E$, especially for transitions with large $\Delta E$,  the rate coefficients for Mg are generally larger than those for Li and Na, though converging at small $\Delta E$.  The explanation for this involves a number of factors, but the main reasons seem to be three-fold, and can be illustrated by considering the transition from the ground to the first excited state.  

First, due to the different atomic structure of Mg, there is a tendency for analogous transitions to have larger $\Delta E$ in Mg than in Li and Na; the ground and first excited states are separated by 2.7 eV in Mg rather than around 2 eV in Li and Na.  Second, the larger ionisation potential of Mg compared to Li and Na means that avoided ionic crossings generally occur at shorter internuclear distances and therefore the radial coupling between states is stronger. Though the crossing occuring at smaller internuclear distance generally decreases the cross-section, the increase in coupling more than compensates for this.  In the case of the radial coupling between the ground and first excited state of MgH, the coupling peaks at around an internuclear distance of 5 a.u.\ with a magnitude of around 0.4 a.u.\ \citep{2010CPL...488..145G}, while Li and Na both have peaks at around 7 a.u.\ with magnitudes of around 0.25 a.u. \citep{1993JPhB...26...61G,2003PhRvA..68f2703B,1985PhRvA..31.3977M,2010PhRvA..81c2706B}.   This latter effect in particular leads to generally larger cross-sections between low-lying states.   Finally, the statistical weights of the entrance channels in the important molecular symmetry ($^2\Sigma^+$ in Mg, $^1\Sigma^+$  in Li and Na) are generally larger in Mg than in Li and Na.  For the case of the initial channel being the ground state, in Mg the $^2\Sigma^+$ symmetry is the only state available, and thus the statistical weight is unity.  However, in Li and Na, the statistical weight of the initial channel is only 0.25.  These factors combined lead to a cross-section for excitation from the ground to first excited state near the threshold being some two orders of magnitude larger for Mg than for Li and Na.  This effect is reduced in the rate coefficient due to the larger $\Delta E$, but still leads to a rate coefficient roughly an order of magnitude larger.    
 
One can also see in  Fig.~\ref{fig:rates} that the rates for charge transfer processes are relatively large in some cases compared to those for excitation, especially from the $4s \: ^1\mathrm{S}$ state.  As in Li and Na, the first excited $S$-state again provides the largest charge transfer cross-section and rate coefficient.  The rate coefficient for this process at 6000~K is $3.20 \times 10^{-9}$~cm$^3$~s$^{-1}$ and is slightly larger, though of the same order of magnitude, than those from the first excited $S$-states in Li and Na, $1.72 \times 10^{-9}$~cm$^3$~s$^{-1}$ and $2.03 \times 10^{-9}$~cm$^3$~s$^{-1}$, respectively.  The reasons that the first excited $S$-state generally provides the largest rate coefficients for ion-pair production are that, in these three cases, the first excited states lead to ionic crossings at intermediate internuclear distances where the transition probability becomes optimal (for details see discussion under \S III.C.2, ``The second mechanism'', in \citealt{mgh_pra}) and that $S$-states lead to the largest statistical weights for the initial channels.

As mentioned, in non-LTE calculations for Li and Na, excitation processes were found to be less important than charge transfer processes; the only process with any appreciable effect was the charge transfer involving the first excited $S$-state.  It is notoriously difficult to predict effects in non-LTE models, and this case is complicated by the fact that Mg, unlike Li and Na, has different spin terms, singlet and triplet, with large collisional rates which are only weakly radiatively coupled. This fact, together with the generally higher rates for excitation processes, might lead one to expect that collisional excitation could be important.  However, any conclusion must await detailed non-LTE calculations.

\begin{acknowledgements}
This work was supported by the CNRS national programme Physique Stellaire (PNPS), the Royal Swedish Academy of Sciences, the Wenner-Gren Foundation, G{\"o}ran Gustafssons Stiftelse and the Swedish Research Council.  P.S.B is a Royal Swedish Academy of Sciences Research Fellow supported by a grant from the Knut and Alice Wallenberg Foundation.  P.S.B. is grateful for the hospitality of the Research School of Astronomy and Astrophysics at the Australian National University where parts of this paper were written.  A.K.B. also gratefully acknowledges the support from the Russian Foundation for Basic Research (Grant No. 10-03-00807-a) and from the MSME laboratory (Universit\'e Paris-Est and UMR 8208 of the CNRS).  The authors acknowledge the role of the SAM collaboration (http://www.anst.uu.se/ulhei450/GaiaSAM/ ) in stimulating this research through regular workshops.
\end{acknowledgements}

\bibliographystyle{aa} 
\bibliography{mgh}

\end{document}